\documentclass[]{spie}  
 \usepackage{pbox}
\usepackage{multirow}
\usepackage{amsmath,amsfonts,amssymb}
\usepackage{graphicx}
\usepackage{floatrow}
\usepackage[export]{adjustbox}
\graphicspath{{./figures/}} 
\usepackage{subcaption}
\usepackage{sidecap}
\usepackage[colorlinks=true, allcolors=blue]{hyperref}
\usepackage{enumerate}
     \usepackage[flushleft]{threeparttable}
     \usepackage{array}
\usepackage{booktabs}
\usepackage [english]{babel}
\usepackage [autostyle, english = american]{csquotes}
\MakeOuterQuote{"}

\title{Panoramic SETI: on-sky results from prototype telescopes and instrumental design}

\author[a]{J\'er\^ome Maire}
\author[a,b]{Shelley A. Wright}
\author[c,d]{Dan Werthimer}
\author[e]{Franklin P. Antonio}
\author[a]{Aaron Brown}
\author[f]{Paul Horowitz}
\author[d]{Ryan Lee}
\author[c,g]{Wei Liu}
\author[h]{Rick Raffanti}
\author[a,b]{James Wiley}
\author[a,b]{Maren Cosens}
\author[i]{Carolyn M. Heffner}
\author[i]{Andrew W. Howard}
\author[j]{Remington P. S. Stone}
\author[k]{Richard R. Treffers}
\affil[a]{Center for Astrophysics \& Space Sciences, University of California San Diego, USA}
\affil[b]{Department of Physics, University of California San Diego, USA}
\affil[c]{Space Sciences Laboratory, University of California Berkeley, CA, USA}
\affil[d]{Department of Astronomy, University of California Berkeley, CA, USA}
\affil[e]{Qualcomm, San Diego, CA, USA}
\affil[f]{Department of Physics, Harvard University, Cambridge, MA, USA}
\affil[g]{Institute of RF- \& OE-ICs, Southeast University, Nanjing, Jiangsu, China}
\affil[h]{Techne Instruments, Oakland, CA, USA}
\affil[i]{Astronomy Department, California Institute of Technology, Pasadena, CA, USA}
\affil[j]{University of California Observatories, Lick Observatory, USA}
\affil[k]{Starman Systems, Alamo, CA, USA}

\authorinfo{Further author information: (Send correspondence to J.M.)\\J.M.: E-mail: jmaire@ucsd.edu}

\begin{document} 
\maketitle

\begin{abstract}
The Panoramic SETI (Search for Extraterrestrial Intelligence) experiment (PANOSETI) aims to detect and quantify optical transients from nanosecond to second precision over a large field-of-view ($\sim$4,450\,square-degrees). To meet these challenging timing and wide-field requirements, the PANOSETI experiment will use two assemblies of $\sim$45 telescopes  to reject spurious signals by coincidence detection, each one comprising custom-made fast photon-counting hardware combined with  ($f/1.32$) focusing optics. Preliminary on-sky results from pairs of PANOSETI prototype telescopes (100\,sq.deg.) are presented in terms of instrument performance and false alarm rates. We found that a separation of $>$1\,km between telescopes surveying the same field-of-view significantly reduces the number of false positives due to nearby sources (e.g., Cherenkov showers) in comparison to a side-by-side configuration of telescopes. Design considerations on the all-sky PANOSETI instrument and expected field-of-views are reported.
\end{abstract}


\section{INTRODUCTION}

High-speed time-resolution astrophysics offers the unique opportunity to study extreme environments of diverse objects such as cataclysmic variables, pulsars, X-ray binaries, and stellar pulsations\cite{Shearer2010}. Some of the most energetic and unusual phenomena in the universe, such as high-speed collisions related to supernovae, blazars, and gamma-ray bursts, can be sources of gamma-rays
and cosmic rays which, when hitting Earth's atmosphere, produce very brief flashes of Cherenkov radiation generated by
the cascade of relativistic charged particles\cite{Hill1961}. The Cherenkov radiation arrives
at the ground in a flash of only a few nanoseconds duration\cite{Park2015}
and can thus be separated from the night-sky background.
Earth's atmosphere is used as the detection medium by Imaging Air Cherenkov Telescopes such as the Whipple telescope\cite{Whipple2007}, 
H.E.S.S.\cite{HESS2004}, MAGIC\cite{MAGIC2007}, FACT\cite{FACT2007}, VERITAS\cite{VERITAS2009} and MACE\cite{MACE2017} which have demonstrated the potential and maturity of this detection technique.

High-time resolution instruments are of interest 
to search for technosignatures by means of detecting nano- to milli-second light pulses that could have be emitted, for instance, for the purpose of interstellar communications or energy transfer. Several programs aimed to search for technosignatures using optical  wavelengths \cite{Werthimer2001, Horowitz2001, Reines2002, Howard2004, Stone2005,  Tellis2017} including near-infrared\cite{Maire2019} have been performed.
The first optical SETI all-sky surveys\cite{Howard2000, Howard2007} with 0.32\,sq.deg of instantaneous field-of-view adopted a transit observing strategy to cover the sky in 150 clear nights. However,  assemblies of single-aperture  telescopes  capable  of  observing  different  parts  of  the  sky instantaneously are  still  needed  to  survey  the  entire sky efficiently and continuously. 

The Pulsed All-sky Near-infrared Optical SETI (PANOSETI) experiment\cite{Maire2018,Wright2018,Cosens2018,Wright2019} is an all-sky observatory project aiming to detect transients that will cover a wide range of timescales in search for nano- to second pulsed light signals, across all optical wavelengths. Based upon two assemblies of $\sim$45 0.46-m Fresnel-lens  telescopes 
equipped with fast low-noise MMPC (Multi-Pixel Photon Counter) detectors operating in the 0.32--0.85\,$\mu$m spectral range, the PANOSETI instrument provides sufficient sensitivity to detect petawatt pulsed signals that could have been sent from kiloparsec distances and beamed toward our direction, that would be distinguishable from most known astrophysical sources from our perspective. PANOSETI will be capable of detecting gamma-rays and cosmic rays with energy above some tens of TeV.
Each part of the sky will be observed simultaneously from two locations for direct detection and confirmation of transients.

We describe PANOSETI telescopes and their required calibrations in Sec.\,\ref{sec2}. The deployment of two pairs of telescopes at Lick Observatory and Palomar Observatory, with the use of a 1-km baseline, is described in the following sections along with obtained results (Sect.\ref{sec2}-\ref{sec5}). Considerations of the production PANOSETI experiment are reported in Sec.\,\ref{sec6}.

\section{PANOSETI module}\label{sec2}

Affordable, lightweight, refractive Fresnel lenses have been used for the detection of nanosecond optical showers generated by high-energy cosmic-rays striking the top of the Earth's atmosphere \cite{Bunner1967,Bunner1968,Stephan2013,Fujii2018}. Even though these lenses have moderate angular resolution (a few arcminutes\cite{Maire2018}), their relatively large-collecting apertures ($>$0.4m) and small focal ratios makes them optimal for a low-angular resolution wide-field survey.
Each PANOSETI module (or "telescope") focuses the incoming light using a 0.46-m $f/1.32$ Fresnel lens (Orafol SC214) built with concentric 0.5-mm equal-width grooves which replace the curved surface of a conventional optical lens, acting as individual refracting surfaces, and bending parallel light rays to a common focal point.
The lens is made of clear optical 1.8-mm thick acrylic material, with a high transmittance in the optical ($>96\%$ in the 0.3--1.6$\mu$m in visible and near-infrared bands) with an anti-reflective coating applied to the lens surfaces  improving transmittance by $\sim2\,\%$. The lens frame has been designed to accommodate thermal expansion\cite{Cosens2018} and includes a 2-mm thick clear acrylic plate to protect the grooved part of the lens against dust and condensation. 
The small focal ratio of the Fresnel lens allows formation of  wide field-of-view (9.97$\times$9.97\,$^\circ$) images of distant objects at the focal plane of the instrument where detectors are located. The telescope Point-Spread Function (PSF) is slightly under-sampled on the optical axis, one-fourth of a pixel full-width-half-maximum (FWHM), and more extended off-axis (up to $\sim$1 pixel FWHM in the corner of the field-of-view).



Three  baffles of graduated sizes are attached along the inside of the telescope tube frame  to limit reflections from the inner surfaces of the telescope\cite{Brown2020} and to block stray light resulting from the faceted and discontinuous surface of the lens. The shapes of the baffles are designed using superellipses, i.e., intermediate shapes between the circular shape of the lens and the square shape of the detector array.
The telescope assembly, tube, frame, and baffles are painted in a black matte finish to further absorb unwanted light\cite{Brown2020}. 
The prototype telescopes are oriented using altitude-azimuthal mounts. Once set, their pointing directions are fixed during an observation night.


The telescope's 32x32-pixel detector array is made of four adjacent  detectors, each one subdivided into four adjacent 8x8-pixel MMPC (Multi-Pixel Photon Counter, Hamamatsu S13361-3050AE-08) detector arrays  which operate in the spectral range 0.32 to 0.85\,$\mu$m with a peak sensitivity at 0.45\,$\mu$m wavelength. These silicon photomultipliers (SiPM) are comprised of Geiger-mode-operated avalanche photodiodes with a high internal gain to enable single photon detection while featuring low dark count ($<$1\,Mcps), high photon detection efficiency (45$\%$), and excellent timing resolution. Each 3mm pixel is made of 50\,$\mu$m micro-cells detecting photons identically
and independently. The sum of the discharge currents from each of these
individual binary micro-cells combines to form a summed photon output, 
thus giving information on the magnitude of an incident photon
flux.
Each of the $1,024$ detector signals is amplified, pulse shaped, and peak detected using an application-specific integrated circuit (ASIC) WEEROC-MAROC3A read-out chips\cite{Barrillon2006,Blin2010} controlled by  FPGAs (field-programmable gate arrays) also used to modify the detector setup, to timestamp frames, and send data over a 1\,GB fiber-optics communications system to a 10\,GB network switch, itself relaying data from all telescopes to the central computer\cite{Liu2020}.
The timing synchronization between detectors and telescopes is maintained by a fiber-connected White Rabbit system\cite{Moreira2009} providing nanosecond synchronization accuracy, and absolute timing through the use of a GPS disciplined time-frequency reference\cite{Liu2020}.

Highly-stabilized power-supplies have been specifically designed for PANOSETI in order to bias detectors as well as to power electronic boards,  activate the internal pulser light source and mechanical parts of the telescopes, such as the focus stage and detector shutter.

In order to examine pulse widths over a large range of time scales (nanosecond to second), the instrument has two observing modes that can be run simultaneously, each one covering a different pulse width range. For detecting pulses shorter than the amplifier shaping time ($<$ 200\,ns), the instrument measures the pulse height of the signal at each pixel ("PH mode"). For time scales larger than the amplifier shaping time, the instrument counts the number of photons in each pixel ("imaging mode") at a programmable frame rate end exposure time. This mode is useful for detecting transients with pulse widths larger than 10$\mu$s, as well as for detector  calibrations.


The discriminator thresholds (DAC) applied to the pulse height detection can be modified to change the sensitivity of the detectors to a given minimal number of photo-electrons (p.e.) per pulse. Fig.\,\ref{fig:sub1}-a shows the number of pulses obtained under dark conditions as a function of the DAC discriminator threshold for two different pre-gain settings in both imaging and PH modes. The steps in the curves  are used to establish a relationship between DAC value and p.e.\ thresholds. In PH mode, the ADC intensity of each channel is also characterized  from the identification of p.e.\ steps in triggered frames at different DAC thresholds (Fig.\,\ref{fig:sub1}-b).

\begin{figure}[hbt]
\centering
 \includegraphics[width=.47\linewidth]{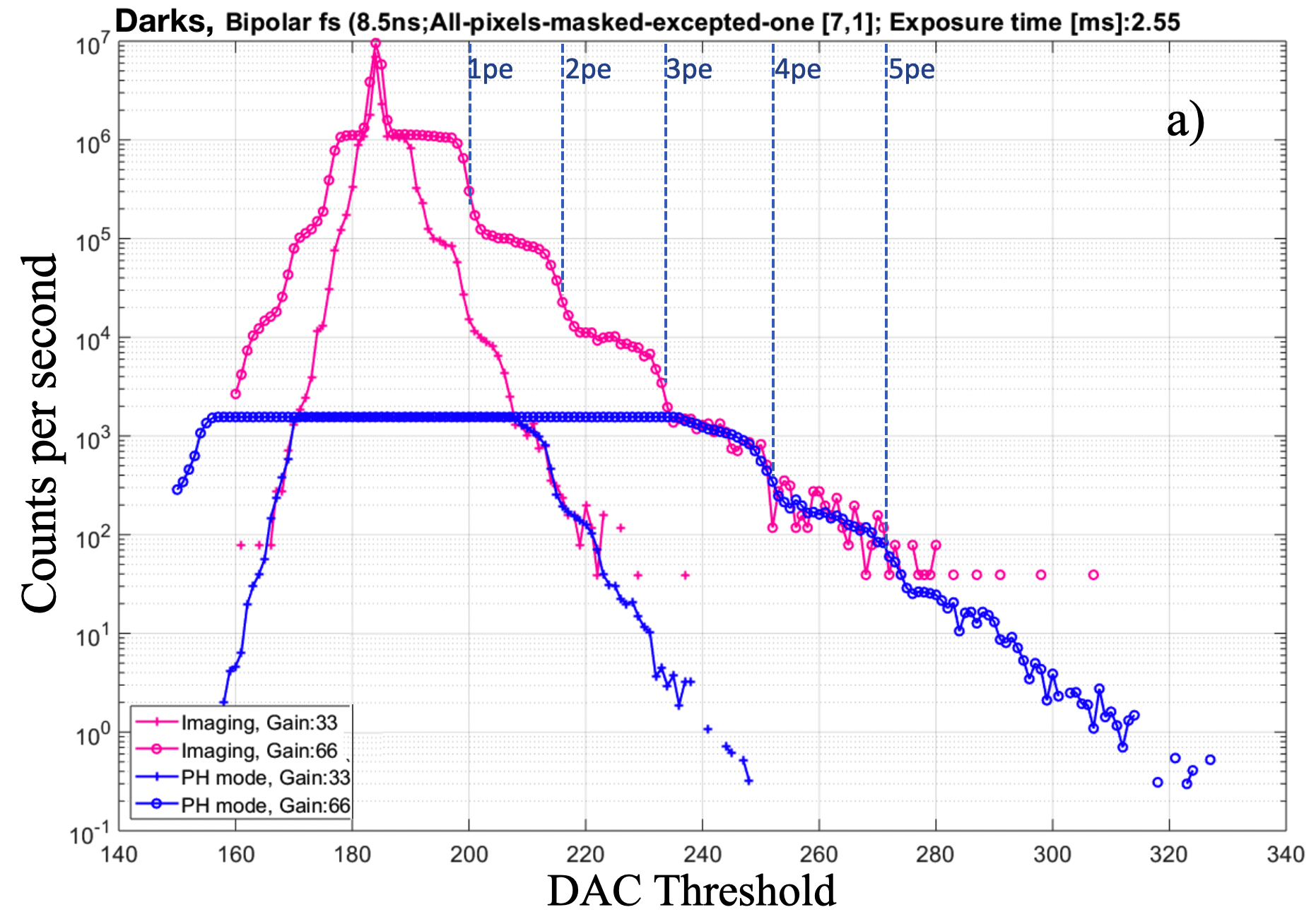}
  \includegraphics[width=.48\linewidth]{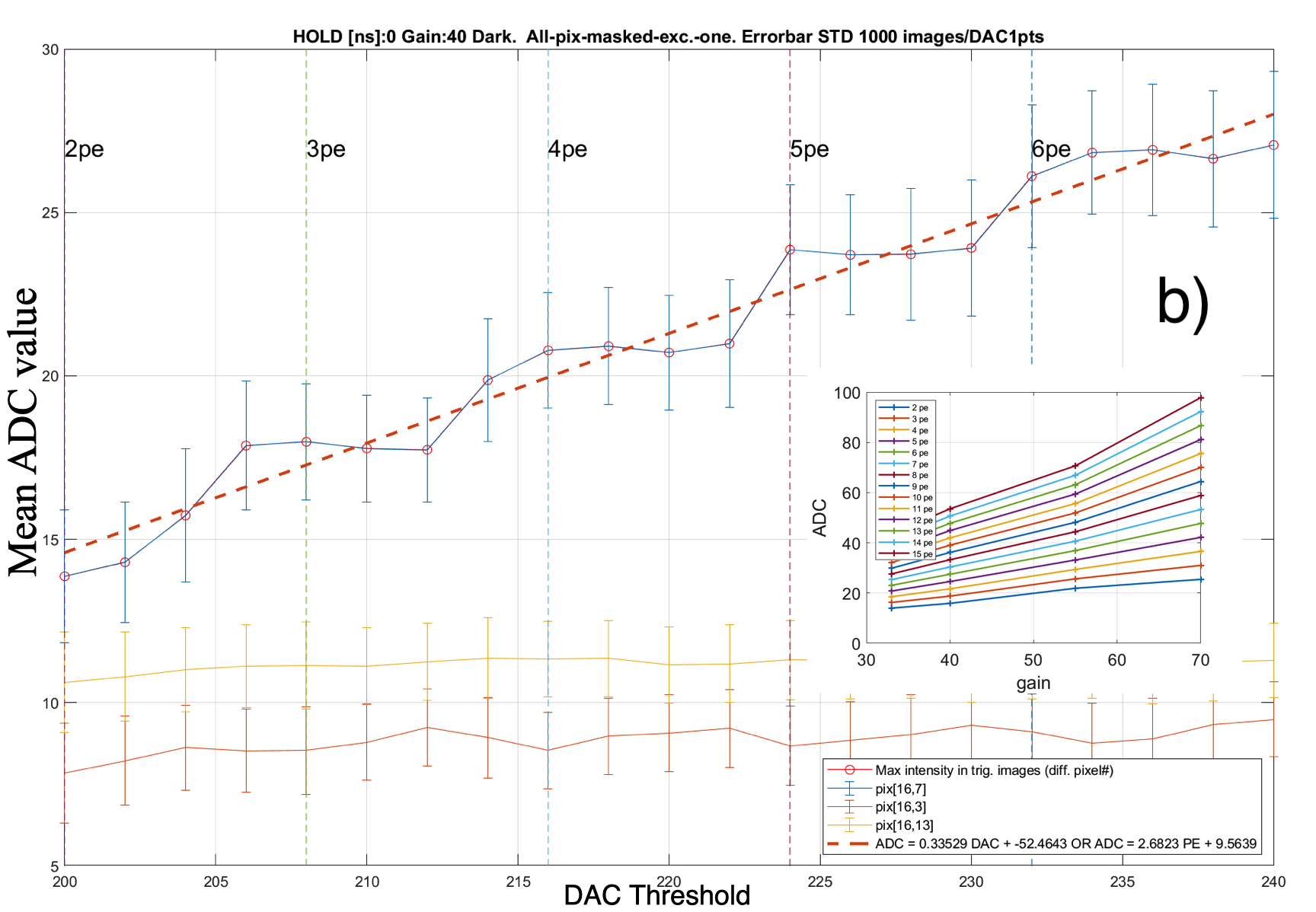}
    \includegraphics[width=.2\linewidth]{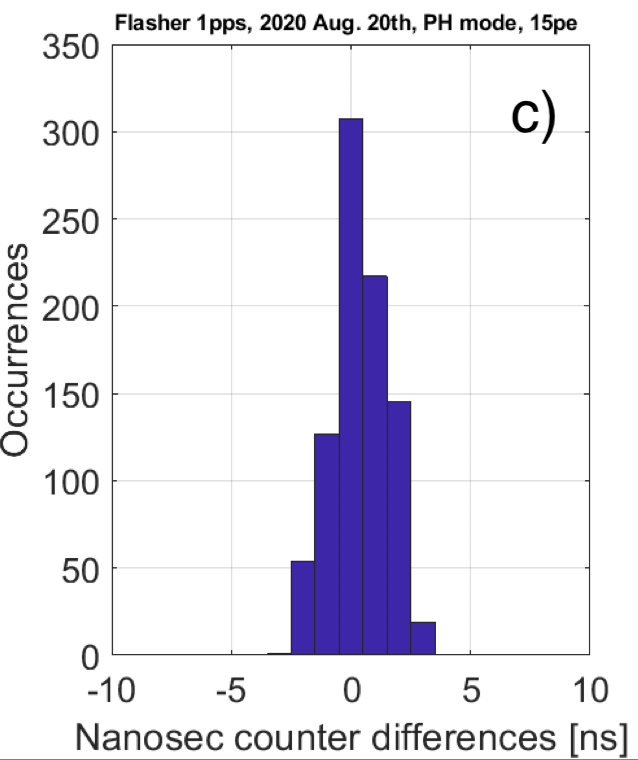}
    \includegraphics[width=.37\linewidth]{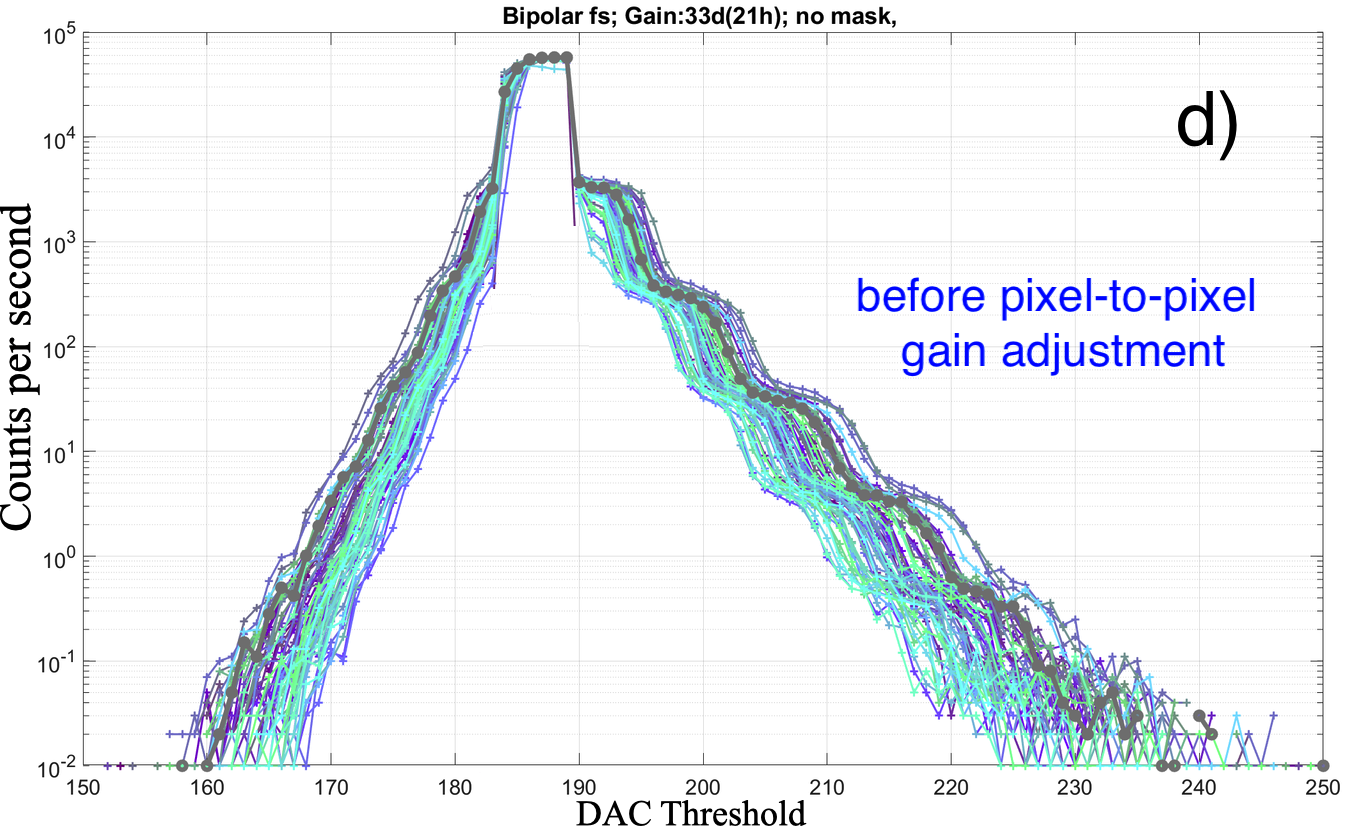}
  \includegraphics[width=.37\linewidth]{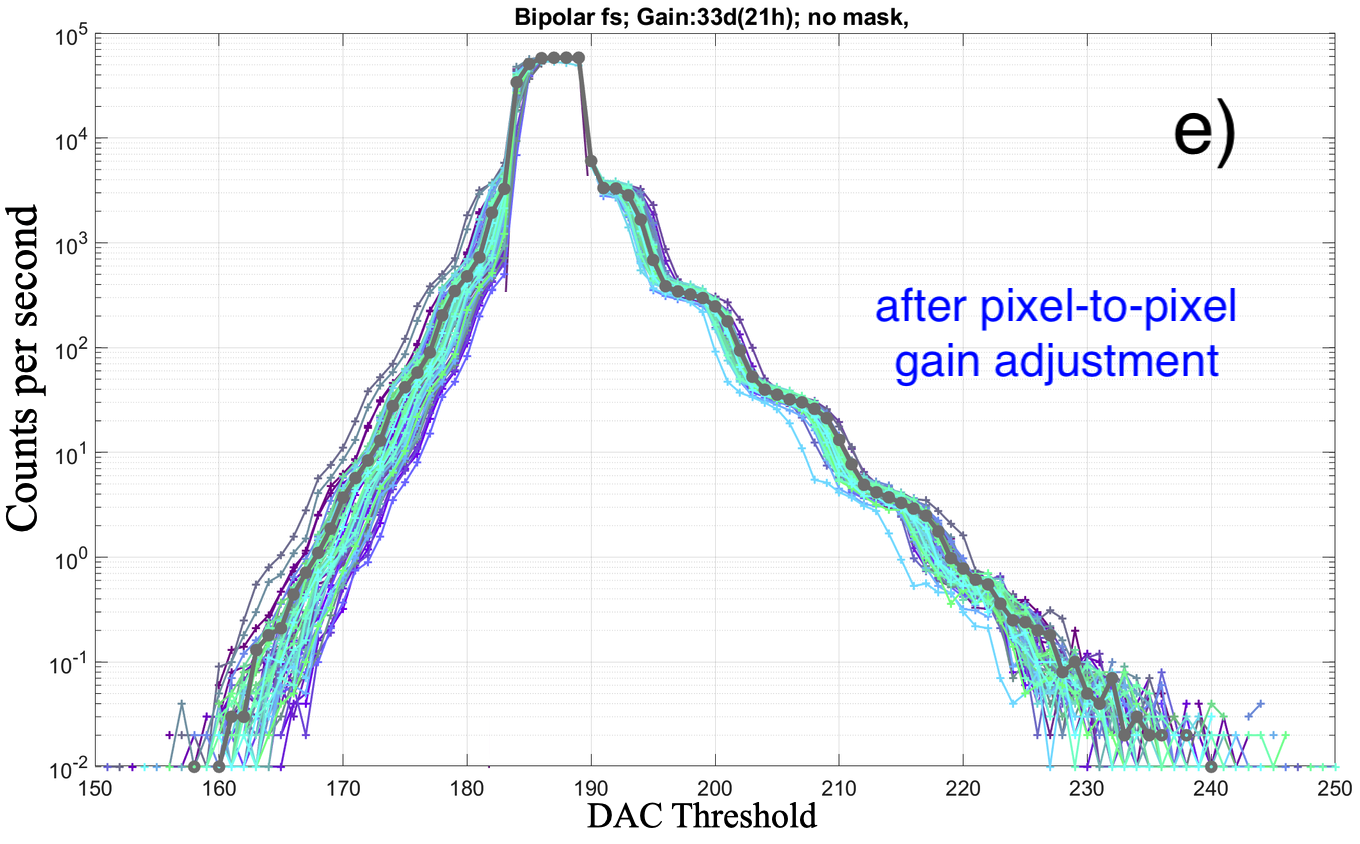}
  \caption{PANOSETI detector calibrations: panel (a) represents the number of pulses counted per second as a function of the DAC threshold for a given detector channel in both imaging and pulse height modes at two different gain settings. Individual photo-event (p.e.) levels can be identified and are used to calibrate the DAC threshold setting. Sweeping the DAC threshold in PH mode is used to calibrate the ADC pulse height output (panel b) as p.e.\ levels can also be identified in triggered ADC values. Panel (c): Histogram of time of arrival difference while observing the flasher source internal unit to verify the nanosecond timing synchronization between telescopes. (d) and (e) panels: number of pulses as a function of DAC threshold for all 64 pixels of a detector quadrant, before and after gain adjustment, showing the equalization of pixel sensitivity at a given threshold. }
  \label{fig:sub4}
\end{figure}

Since the same discriminator threshold value is applied to each different pixel of a 8x8-pixel array, the sensitivity of each individual channel is slightly different and is therefore adjusted. The variable gain pre-amplifier of the MAROC3A allows the gain of each channel to be tuned and is able to compensate the non-uniformity between detector channels. Fig.\,\ref{fig:sub1}-d-e  show the pixel-to-pixel variations of sensitivity before and after performing a pre-amplifier compensation where  p.e.\ steps have been aligned to the same threshold values.
In addition to telemetry metadata which includes detector currents and voltages, the instrument monitors each detector's temperature to adjust the optimal detector biases (with adjustments of 54mV/$^\circ$C ) in real-time to maintain a constant detector gain.

Coincident detection between two or more sites requires precision time stamping of events at each site. Prior to observations, a light source unit is used to send flashes on both detectors at various repetition rates  to check the timing synchronization between telescopes. The light pulses from the flasher are sent through fibers of equal length to both telescopes. A histogram of measured time of arrival differences between telescopes are represented on Fig.\,\ref{fig:sub1}-c showing a $\pm$\,2\,ns accuracy in timing synchronization.

For each telescope pointing an astrometric solution is determined using on-sky imaging-mode observations of bright stars (V$<$4). Since the telescope pointings are fixed in time it is convenient to obtain the altitude-azimuth coordinates for each pixel. Known stellar coordinates are used during the time of observations to give astrometric reference points before performing a gnomic projection to obtain the coordinates of each pixel centers\cite{Calabretta2002,Greisen2002}. Observations of several stars at different orientations in the field-of-view are used to generate a distortion solution and plate constants. The "distortion map" is then used to correct the pixel coordinates with respect to the gnomic projection. These astrometric calibrations are performed for each telescope and are then used to determine residual misalignment offsets between each telescope's field orientation on-sky.

\section{Observations} \label{sec3}
\subsection{Lick Panograph Experiment}
 Two identical PANOSETI telescopes have been deployed side-by-side in the  Astrograph dome\cite{Osterbrock2007} at Lick Observatory (Fig.\,\ref{fig:sub1}-left) since January 2020 to search for transient signals. It also has been used as a test bed instrument to verify instrumental performance and to characterize false alarm rates. The pointing directions of the two telescopes were set to be identical to provide direct confirmation of detected transient signals. Observations with the PANOSETI telescopes in the Astrograph, coined "Panograph" instrument, are performed remotely which allows our team to observe and test new software as the project develops.  
 
\subsection{Palomar baseline experiment}
\begin{figure}[htb]
\centering
  \includegraphics[width=.45\linewidth]{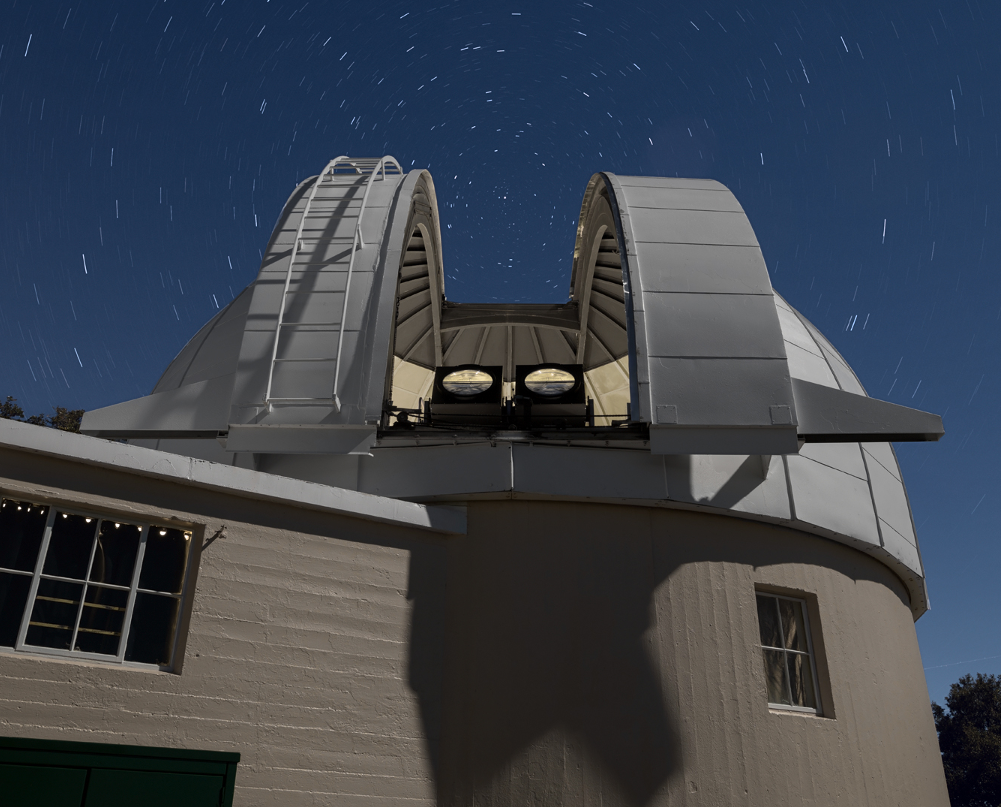}
\includegraphics[width=.515\linewidth]{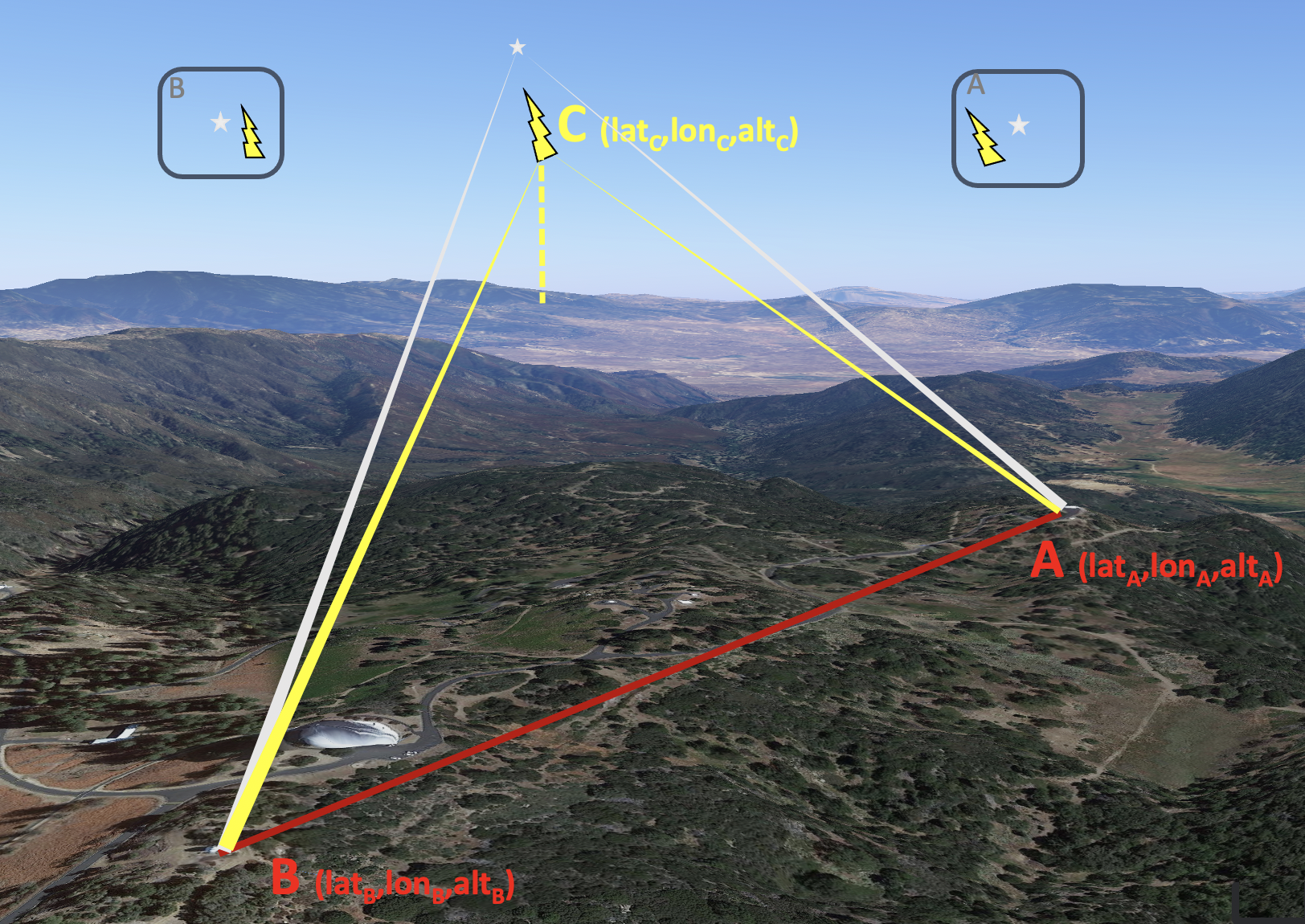}
  \caption{LEFT: Two side-by-side PANOSETI telescopes pointing in the same direction have been deployed inside the Astrograph dome at Lick Observatory (photo courtesy: Laurie Hatch Photography). RIGHT: 1-km baseline experiment at Palomar Observatory, the two telescopes are located on sites A and B and pointing at the same star. This baseline introduces a measurable parallax angle for flashes of light generated at finite distances in Earth's atmosphere and a time of arrival difference offset with respect to astronomical sources.}
  \label{fig:sub1}
\end{figure}

In order to verify the expected number of false alarms generated by instrumental noise and Cherenkov showers (see Sec.\ref{sec5} below), successive experiments using two telescopes separated by 250-m and 1-km baselines were performed at Palomar Observatory on Aug.\,21$\mathrm{^{st}}$  and Oct.\,9$\mathrm{^{th}}$ 2020, respectively. The long baseline separation introduces a measurable parallax angle (illustrated on Fig.\,\ref{fig:sub1}, right-panel) for flashes of light generated in Earth's atmosphere and a time arrival offset compared to astronomical sources (see Sect.\ref{sec4}). The two telescopes were deployed in the field using a long 1.2km duplex fiber for timing synchronization and data transfer. Co-alignment of telescope pointing direction was performed by adjusting pointing directions while  observing in real-time bright planets (Jupiter and Mars) in imaging mode.

\section{Experimental design}\label{sec4}

\subsection{Time of Arrival difference}
\begin{figure}[htb]
\centering
  \includegraphics[width=0.95\linewidth]{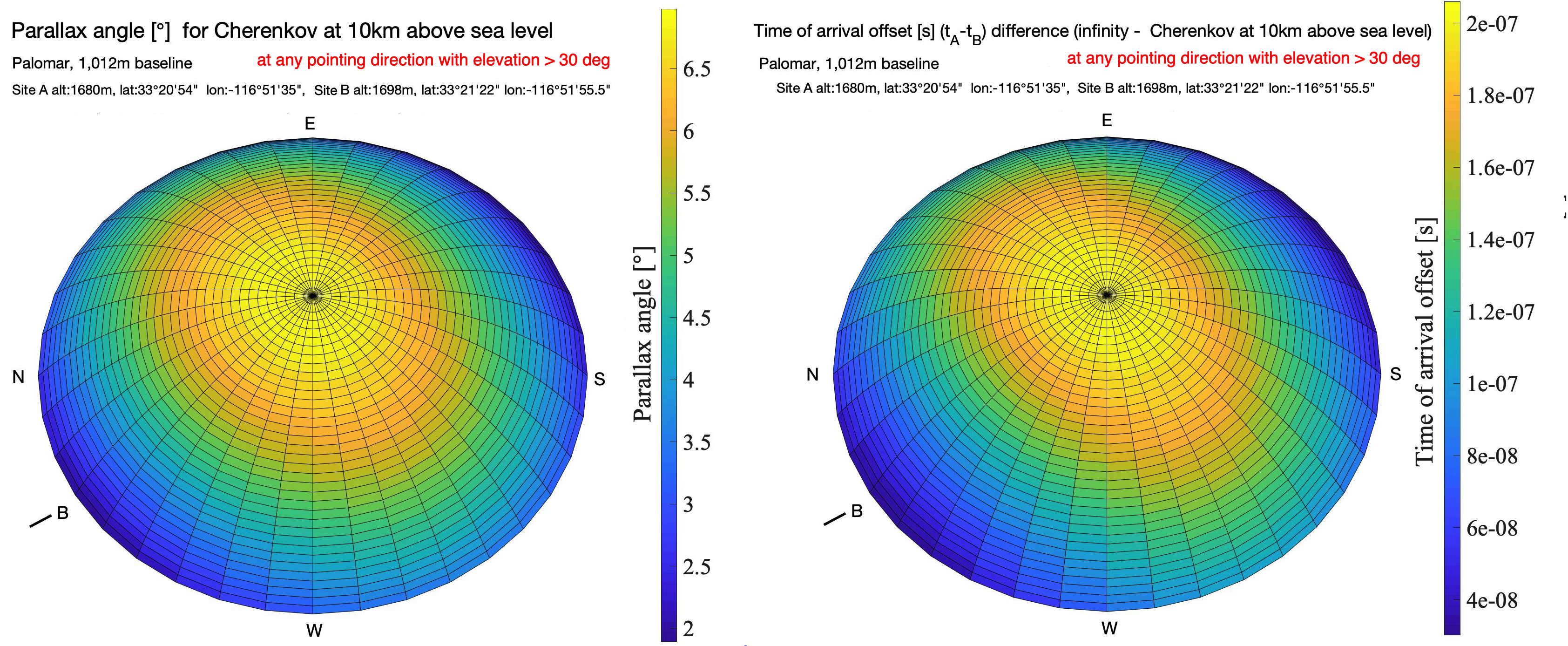}
  \caption{LEFT: Expected parallax angle of a Cherenkov shower occurring at 10\,km above sea level for a 1\,km PANOSETI baseline in any pointing direction above 30$^\circ$ elevation. RIGHT: Expected difference between time of arrival difference for a source at infinity and a phenomena occurring 10\,km above sea level (e.g., Cherenkov shower maximum) using the same 1\,km Palomar baseline, in any pointing direction above 30$^\circ$ elevation.}
  \label{fig:sub3}
\end{figure}

 Accurate time stamping of detected events at multiple sites enables the ability to measure the distance of nearby phenomena in Earth's atmosphere or low-Earth orbit, e.g., false alarms from atmospheric Cherenkov radiation or optical transients from satellite glints.
For a given baseline AB (illustrated on Fig.\,\ref{fig:sub1}-right), the time of arrival difference between two sites depends on the pointing direction of each telescope and distance of the detected event between the two sites. For any event C occurring at coordinates $(\mathrm{lat}_C,\mathrm{lon}_C,\mathrm{alt}_C)$ 
with respect to the World Geodetic System of 1984 (WGS84), one can calculate the distances CA and CB of a event from each telescope considering the known WGS84 coordinates of sites A and B. We then determine the optical path length difference by considering the change of refractive index of air along the lines of sight CA and CB given the changes in pressure and temperature with altitude with respect to the 1976 Standard Atmosphere. We used the Ciddor method\cite{Ciddor1996} to calculate the change of refractive index of air along the path as a function of pressure and temperature (for a 0.5\,$\mu$m wavelength, 50$\%$ humidity and a CO$_2$ concentration of 450 ppm). The expected time of arrival difference $\Delta t= t_A - t_B$ is then determined from the calculated optical path length difference and speed of light in that medium. The time of arrival difference is negligible for side-by-side telescopes but increases with the baseline distance.
If the baseline is long enough, the difference of $\Delta t$ between a nearby source and a distant astronomical sources in any pointing directions becomes measurable considering the timing accuracy of the instrument.
Fig.\,\ref{fig:sub3}-right represents the difference of $\Delta t$ between a phenomena occurring at 10km above sea level (typical Cherenkov shower maximal altitude is 12.8\~km \cite{Weekes2003}) and a source at infinity for a 1km baseline at Palomar Observatory in any pointing directions above 30$^\circ$ elevation. The minimum of this difference (37\,ns) occurs at lower elevations for pointing directions along the baseline direction and a maximum  of $\sim$200\,ns close to zenith.





\subsection{Coincidence detection}

A time coincidence occurs when both telescopes pointing in the same direction detect an event above our sensitivity threshold (e.g., above 15\,p.e.) during a time interval $\Delta t_{\mathrm{window}}$. To consider the delays in time of arrival due the geometrical orientations of the telescopes, the offset in time of arrival  $\Delta t_{\mathrm{offset}}$ for an object at infinity in a given pointing direction is included, such that a time occurrence occurred when
\begin{equation}
| t_A - t_B - \Delta t_{\mathrm{offset}} | < \Delta t_{\mathrm{window}}
\end{equation}
For distant astronomical sources, $\Delta t_{\mathrm{window}}$ could be chosen such that it will discriminate nearby events from the coincidence detection process. However, the value of $\Delta t_{\mathrm{window}}$ should be large enough to take into account timing accuracy and uncertainty of the instrument ($\sim$5ns at 30$^\circ$ elevation up to $\sim$10ns at zenith for the 1km baseline).

\subsection{Parallax angle}

The Palomar 1\,km baseline yields large parallax angles for optical flashes occurring in Earth's atmosphere that are observed by two separated telescopes. The parallax of such an event is the difference in the apparent positions of the  shower maxima on both telescopes, and is measured by the angle of inclination between those two lines. The parallax angle of an event can be calculated in the WGS84 system as the angle $\Lambda$ between two vectors $\textbf{CA}$ and $\textbf{CB}$, where C represents the location of the shower maximum (see Fig.\,\ref{fig:sub1}) of geodetic coordinates (lat$_\mathrm{C}$,lon$_\mathrm{C}$,alt$_\mathrm{C}$). For any event C, the parallax angle, i.e. the angle  $\Lambda$ between the two vectors \textbf{CA} and \textbf{CB}, can be determined using 

\begin{equation}
    \Lambda= \mathrm{atan2}(|\textbf{CA}\times\textbf{CB}| , \textbf{CA}.\textbf{CB})
\end{equation}

The parallax angle of a shower maximum  can thus be determined at any pointing directions and shower altitudes as shown in Fig.\,\ref{fig:sub3}-left, in the case of the Palomar 1km baseline for a shower occurring at 10km above sea level. The cap represented on the figure represents the pointing directions for any elevation greater than 30$^\circ$. The detected shower maximum as seen by the two separated telescopes will have different coordinates and will be separated by at least $\sim2^\circ$ (6 pixels) at low elevations in the baseline direction up to $\sim7^\circ$ (21 pixels) close to zenith.

\section{Results}\label{sec5}

\begin{figure}[htb]
\centering
  \includegraphics[width=0.57\linewidth]{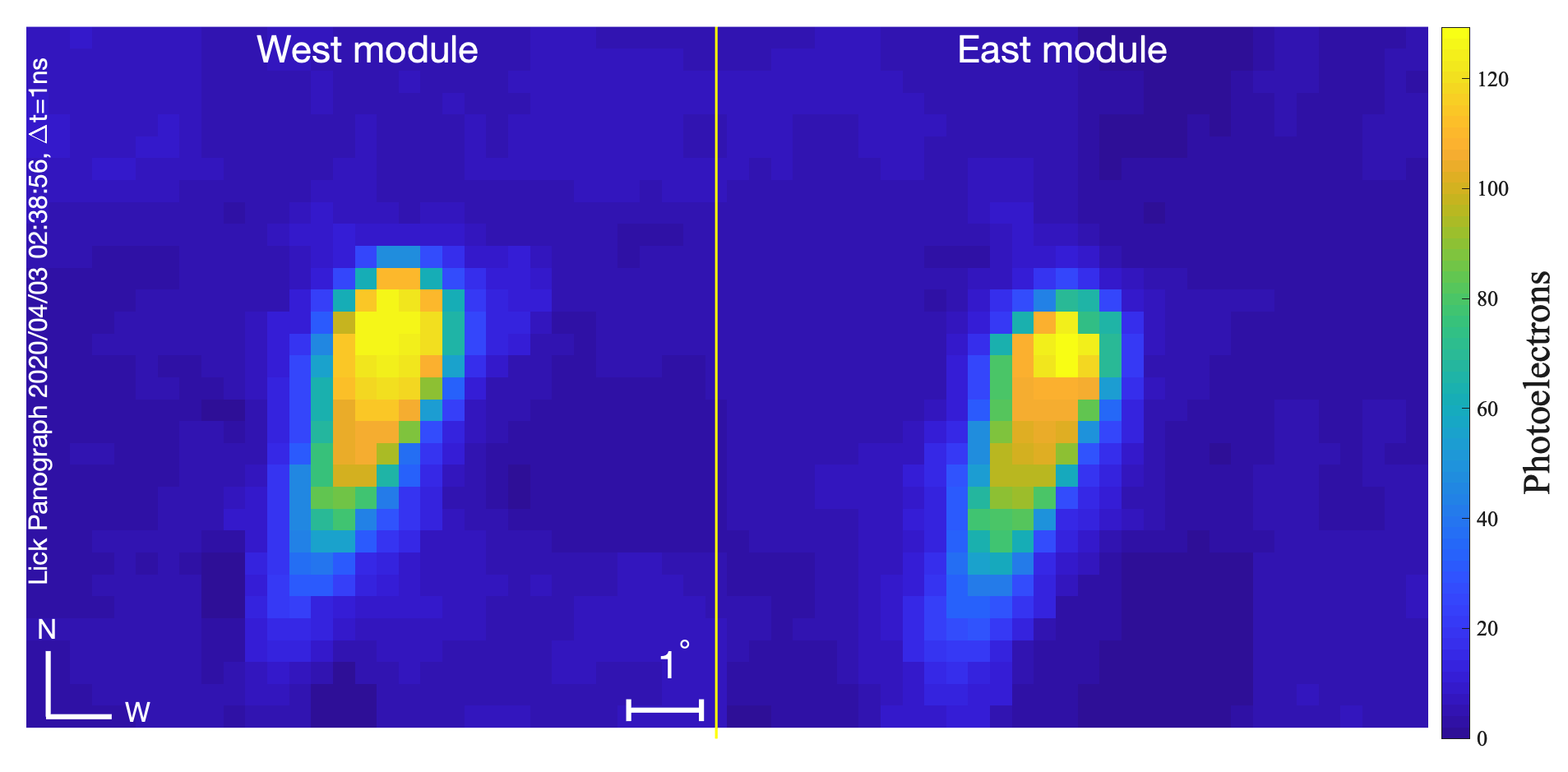}
    \includegraphics[width=0.42\linewidth]{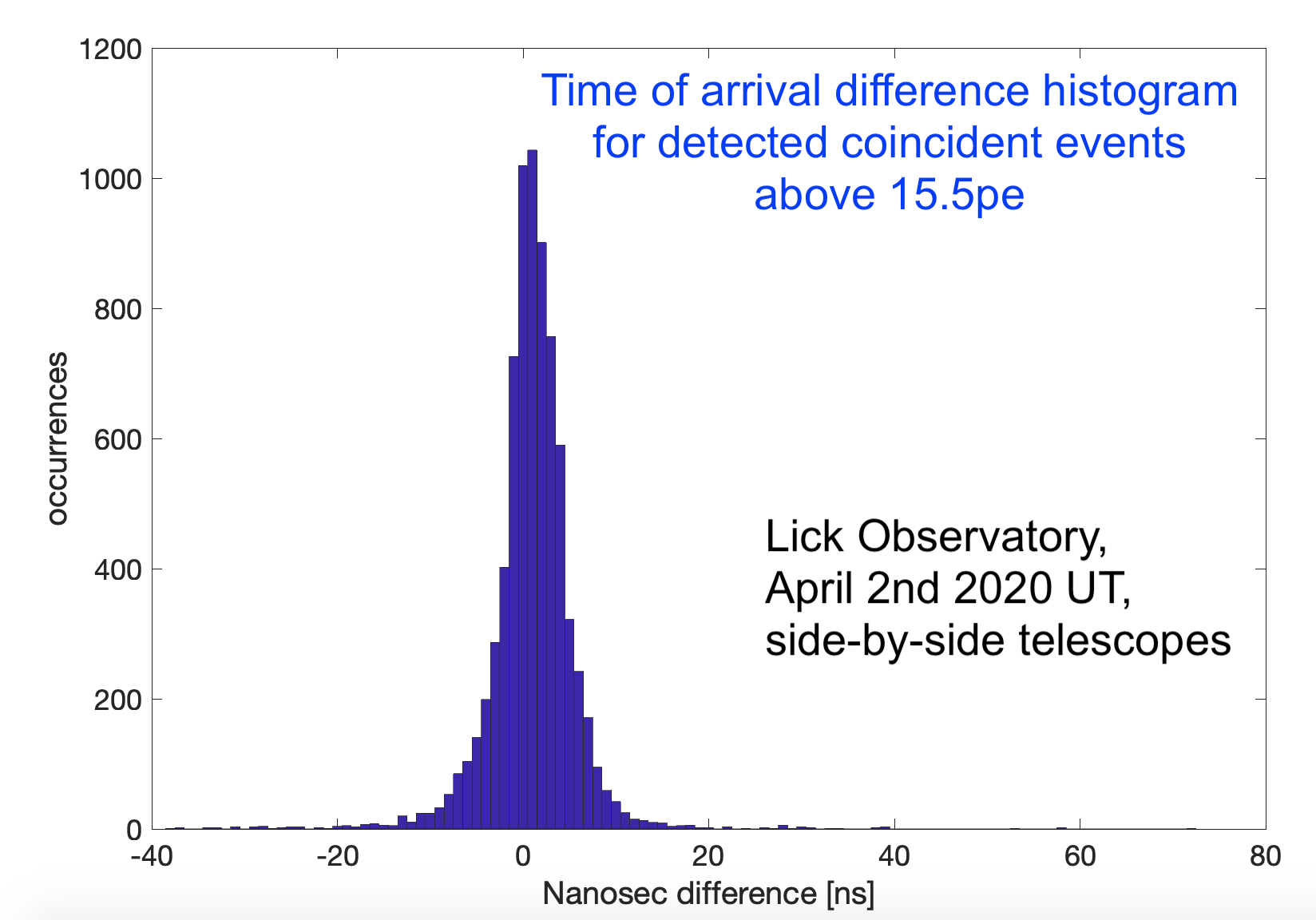}
  \caption{LEFT: An example of a Cherenkov shower detected at Lick Observatory with side-by-side telescopes (slightly misaligned by an offset of 1.5\,pixels). This specific shower extends to several degrees and can be distinguished from a source at infinity. Right: Histogram of time of arrival difference for detected coincidences above 15.5\,p.e.\ at Lick Observatory, on April 2$^{\mathrm{nd}}$ 2020 UT showing that these events are detected within less than 10\,ns difference.}
  \label{fig:sub5}
\end{figure}

We measured the false alarm rate (FAR) at Lick Observatory during a 3\,h period of continuous on-sky observations in PH mode on April 2$^{\mathrm{nd}}$ 2020. Considering coincident events detected by the two telescopes above 15.5\,p.e., we found a false alarm rate of 0.31\,coincidence/s/pair of telescopes (on average 1 coincident event each 3.2\,s per pair of telescopes) with respect to the $9.9^\circ$-wide field-of-view. This 15.5\,p.e.\ threshold was chosen to be above the average  highest threshold  needed to detect pulses generated by sky background (\~5 to 11\,p.e.\ depending on lunar illumination). An example of Cherenkov shower  detected at Lick Observatory is represented on Fig.\,\ref{fig:sub5}. The spread and elongation of the shower over many degrees can be used to discriminate this specific event from an astronomical source. However, in the case of showers developing along the telescope line-of-sight, it is difficult to differentiate them from other point-like source. About 20\,-\,30\,\% of the detected coincidences could not be rejected on criteria based on the shape or spread or angular separation. Since the telescopes are located side-by-side, no rejection based on parallax or time of arrival analysis could be obtained. As can be seen on Fig.\,\ref{fig:sub5}, almost all coincidences have a time of arrival difference between telescopes  included in a $\pm$\,10\,ns interval.

\begin{figure}[htb]
\centering
  \includegraphics[width=1.\linewidth]{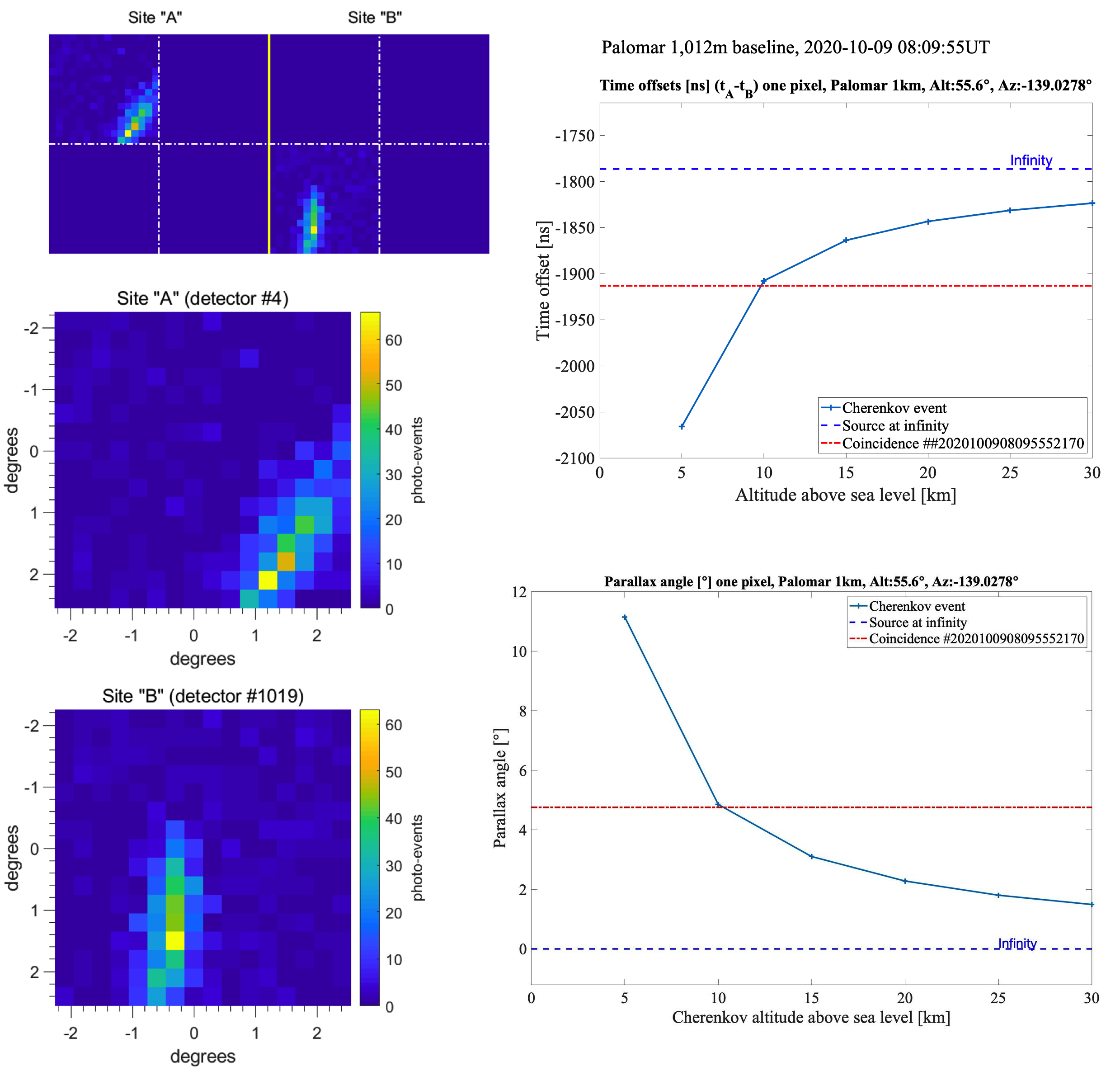}
  \caption{LEFT: Example of a Cherenkov shower detected by the two PANOSETI telescopes at Palomar Observatory using a 1-km baseline on Nov.\,9$^{\mathrm{th}}$, 2020. The images show the triggered events on both detectors and their respective positions on the entire telescope field-of-view (top). The 4.75$^\circ$ offset in position of the shower maximum is due the parallax that the 1km baseline provides for nearby objects. Both the time of arrival difference (1913\,ns) and the parallax (4.75$^\circ$) are in agreement with an event that occurred at about 10\,km above sea level.}
  \label{fig:sub6}
\end{figure}

The false alarm rate was measured at Palomar Observatory (Table \ref{tab:res}) using the same sensitivity threshold with two telescopes separated successively by three different baselines: modules side-by-side, separated by a 250\,m, and by a 1,012\,m baseline. The false alarm rate decreases drastically with baseline distance since the baseline eventually becomes larger than the typical shower detection area ($\sim$\,240-m diameter\cite{Weekes2003}). However, a few rare showers are still detected with the 1-km baseline. We observed 1 event per 18\,min\,45\,s on average, using a large $\Delta t_{\mathrm{window}}$ (200\,ns) at a zenith angle of 35$^\circ$. All detected events are identified as Cherenkov showers since their time of arrival differences and parallaxes are consistent with phenomena occurring between 9 and 12\,km above sea level. Fig.\,\ref{fig:sub6} shows a typical example of these events detected by both telescopes pointing on the same field-of-view with the 1\,km Palomar baseline on October 9$^{\mathrm{th}}$, 2020. Due to the long baseline, the parallax is so large (4.75$^\circ$ at shower maximum) that the event is detected on different quadrants of the two detectors. A 1km baseline decreases the false alarm rate since the apparent angular separation of the detected shower may result in showers falling outside of the second-telescope field-of-view. The time of arrival differences of showers detected with a 1-km baseline are significantly different from an astronomical source (92 to 141\,ns of difference with a source at infinity in their pointing directions).

 \begin{table}[htb]
\caption{Measured PANOSETI false alarm rates (FAR) at Lick Observatory (1,284\,m altitude) and Palomar (1,680\,m\,\&\,1,698\,m altitudes of sites A and B)  considering coincident events detected by the two telescopes above 15.5\,p.e. (without rejection based on parallax or shape of the event). } 
\label{tab:res}
\begin{center}    
\begin{tabular}{|l|l|l|l|l|l|} 
\hline
\rule[-1ex]{0pt}{3.5ex}
& & Eleva- & Obs. & False Alarm Rate   &  False Alarm Rate  \\
Location & Baseline&tion$^a$&Duration&[coincidences/s/pair of telesc.]& [coincid./s/pair of telesc.]\\
&&&&($\Delta t_{\mathrm{window}}$ =200\,ns) & ($\Delta t_{\mathrm{window}} $ =10\,ns)\\
 \hline 
\rule[-1ex]{0pt}{3.5ex}
Lick & Side-by-side & 60$^\circ$ & 3\,h 2\,min & 0.31 (1 coinc. per 3.3\,s) &  0.3 (1 coinc. per 3.2\,s) \\
 \hline 
 \rule[-1ex]{0pt}{3.5ex}
Palomar & Side-by-side &  30$^\circ$& 8\,min\,34\,s  & 0.2 (1 coinc. per 5\,s)$^b$ & 0.2 (1 coinc. per 5\,s)$^b$  \\
 \hline 
 \rule[-1ex]{0pt}{3.5ex}
Palomar & 250\,m & 28$^\circ$& 13\,min\,39\,s  & 0.034 (1 coinc. per 29.3\,s)$^b$ &  0.023 (1\,coinc.\,per\,42.8\,s)$^b$ \\
 \hline 
 \rule[-1ex]{0pt}{3.5ex}
Palomar & 1,012\,m &  55$^\circ$& 1\,h\,15\,min & 8.8 10$^{-4}$ (1\,coinc.\,per\,18\,min\,45\,s) &  0 (no coincidence) \\
 \hline 
\end{tabular}
\end{center}
 \vspace{-0.25cm}\hspace{0.5cm}  a: related to field-of-view center.  b: extrapolated from 3/4 of the  detectors (full detector otherwise).
\end{table} 
 


In summary, these results highlight the importance of the long baseline in excluding false alarms such as Cherenkov showers, as the baseline introduces a large parallax and a timing offset between nearby Earth objects, compared with astronomical sources. The 1-km baseline reduces the number of Cherenkov showers detected by both telescopes due to the angle at which events are detected. Even though a large part ($\sim$70-\,80\,\%) of showers can be identified from their elongated or widely-spread shapes using side-by-side telescopes, the long 1\,km baseline enabled the ability to distinguish all events from astronomical sources at infinity during our spaced-telescope observations. We plan to still use the side-by-side modules for confirmation follow-up and tracking if an interesting candidate source is detected. 

Even though the study of Cherenkov showers is not the PANOSETI primary objective, the long baseline  gives the capability to characterize these events in finer details. Given the different orientations of the elongated shower measured with our data (see for instance Fig.\,\ref{fig:sub6}),  we can use the stereoscopic effect given by the long baseline to deduce the original direction of the cosmic-rays in Earth's atmosphere. For showers induced by cosmic-rays (i.e., electrically-charged particles such as nuclei), their trajectories in space are bent by interstellar magnetic fields making it difficult to identify the true direction from which they originate. However, for particles with the highest energy ($>$10\,EeV, i.e., $>$10$^{19}$eV), their deflection can be small enough that it is indeed possible to identify their sources\cite{Auger2007}.  Unlike cosmic-rays, $\gamma$-ray photons travel in straight lines and are not affected by interstellar magnetic fields, making it possible to deduce the source from which they originate directly from their Cherenkov shower axis. As future work, we plan to determine the energy of the particle or $\gamma$-ray emission that initiated the shower, and, if possible, to identify their sources. All Cherenkov events will be time stamped and recorded in the PANOSETI software system.



\section{Dome assembly}\label{sec6}

Assemblies of high-time-resolution telescopes capable of observing different parts of the sky instantaneously are still needed to survey the entire sky quickly and repeatedly. We presented a conceptual design for a PANOSETI observatory based upon an assembly of refracting $\sim 0.5$\,m Fresnel telescopes tessellating two geodesic domes\cite{Maire2018, Wright2018, Cosens2018}. This design produces a spherical layout of collecting apertures that optimizes the instrument footprint, aperture diameter, instrument sensitivity, and total field-of-view coverage.
While our initial design of the entire assembly had a larger total collecting area, we further optimized it by reducing the tessellation frequency of the geodesic dome structure, and opted for a telescope design that could fit into all triangles of a single-layered geodesic dome. This lower-frequency design significantly minimizes the amount of material required for building the frame structure and telescope support compared to our previous design\cite{Maire2018}. Reducing the tessellation frequency also reduces the number of required customized parts, which gives the capability to use off-the-shelf struts and joints for the geodesic structure that are relatively common and inexpensive. Other considerations, such as the size of available astronomical enclosures capable of protecting the telescope equipment from inclement weather have imposed constraints on the assembly size, telescope apertures, and number of telescopes per assembly. 

The geodesic assembly design proposed for the PANOSETI spatial configuration of telescopes is generated from the triangular subdivision of a truncated spherical icosahedron. Among the five Platonic solids, it has the largest number of identical faces, making it a natural choice for designing an all-sky instrument structure supporting a collection of identical telescopes. Any grid on one face of the spherical icosahedron can be replicated, thus covering the entire sphere with a pattern of regular shapes covering the $\pi$-sr spherical cap that the PANOSETI instrument will be surveying at elevation greater than 30$^\circ$. The tessellation frequency defines the number of segments that the grid divides into the polyhedron edges. 
Various subdivision schemes have been developed to form uniform grids of nearly-identical flat triangles\cite{Popko2012}, differing in the choice of great-circle  arcs to divide the face. 
Each subdivision method gives a grid with different chord lengths, triangle areas, and shapes. We used the \textit{Equal-arcs (three great circles)} subdivision method\cite{Popko2012} to design a uniform grid which has, among the other tested subdivision methods\cite{Maire2018}, the largest diameter of the inscribing circle  that fits into all triangles,  as well as the  smallest variance in strut length and triangle area, and the most uniform distribution of face orientations\cite{Popko2012}.  Telescopes will be attached to the geodesic dome struts using brackets that are adjustable to a few cm irregularities to define the telescope placement with respect to the incenter of each triangle. Attachment points will be added to the telescope tube
 to interface with the geodesic dome at an adjustable orientation\cite{Brown2020}.
 
 \begin{table}[hbt]
\caption{PANOSETI geodesic dome main parameters and effective sky coverage  with redundancy for different configurations of telescope assemblies. } 
\label{tab:onskymapping}
\begin{center}    
\begin{tabular}{|l|l|l|l|} 
\hline
\rule[-1ex]{0pt}{3.5ex}  \textbf{Assembly Configurations} & A & B & C \\
\hline
\rule[-1ex]{0pt}{3.5ex}  \textbf{Dome(s) per observatory} &  1 & 1 & 2\\
\hline
\rule[-1ex]{0pt}{3.5ex}  \textbf{Nb modules per observatory} &  80 & 45 & 90\\
\hline
\rule[-1ex]{0pt}{3.5ex}  \textbf{Tessellation frequency} &  4$^\nu$ & 3$^\nu$ & 3$^\nu$\\
\hline
\multicolumn{4}{l}{\textbf{Geodesic dome sizes:}}\\
\hline
\rule[-1ex]{0pt}{3.5ex} 
\textbf{Dome Diameter (incl. telescopes)  }& 6.2m (20.5ft) &	4.84m (15.8ft) &	Same than 3$^\nu$ \\
\hline
\rule[-1ex]{0pt}{3.5ex} 
\textbf{Dome Height (incl. telescopes)} & 	2.2m (7.3ft) & 1.8m (5.6ft)	 &	Same than 3$^\nu$ \\
\hline
\rule[-1ex]{0pt}{3.5ex} 
\textbf{Possible Enclosures }& AstroHaven 22ft & AstroHaven 18ft & Same than 3$^\nu$ (x2)
\\
\rule[-1ex]{0pt}{3.5ex} 
& LosBerger  10m & LosBerger 6m & 
\\
\rule[-1ex]{0pt}{3.5ex} 
&&
Baader 6.5m 
&  \\	
\hline
\multicolumn{4}{l}{\textbf{Geodesic dome components:}}\\
\hline
\rule[-1ex]{0pt}{3.5ex} 
\textbf{No. of struts}&	130 &	75 &	2x 75\\
\hline
\rule[-1ex]{0pt}{3.5ex} 
\textbf{No. of unique strut length} &	5  & 3 &	Same than 3$^\nu$\\
\hline
\rule[-1ex]{0pt}{3.5ex} 
\textbf{Struts lengths [min - max]}	& 0.92m - 1.07m&	0.93m - 1.07m &	Same than 3$^\nu$\\
\hline
\rule[-1ex]{0pt}{3.5ex} 
\textbf{No. of unique face panels} &	5&	3 &	Same than 3$^\nu$\\
\hline
\rule[-1ex]{0pt}{3.5ex} 
\textbf{Diameter of largest inscribing }&
0.52m (20.4in) &	 0.53m (20.5in) &	Same than 3$^\nu$\\
\textbf{circle which fits in all panels}$^a$ & & &\\
\hline
\rule[-1ex]{0pt}{3.5ex} 
\textbf{No. of joints} &	51 &	31	& 2x 31\\
\hline
\rule[-1ex]{0pt}{3.5ex} 
\textbf{No. of pixels per dome}	& 81,920 &	46,080	 & 92,160\\
\hline
\hline
\rule[-1ex]{0pt}{3.5ex}  \textbf{PANOSETI } &     	7,449 sq.deg. &     	4,441 sq.deg.  & 7,502 sq.deg. \\
\rule[-1ex]{0pt}{3.5ex}   \textbf{instantaneous}   & 	2.27 sr &  		1.35 sr  &  		2.28 sr \\
 \rule[-1ex]{0pt}{3.5ex}  \textbf{field-of-view}    &       18.0\% of the sky &       10.8\% of the sky    &  18.2\% of the sky \\
\hline 
\rule[-1ex]{0pt}{3.5ex}  \textbf{Redundant coverage}  &  422  sq.deg.  &  0 sq.deg.   & 1,442 sq.deg.  \\
\rule[-1ex]{0pt}{3.5ex}  \textbf{Redundant pixels}  &     6.7\%   &     0\%   &   19.2\%  \\
 \hline 
\end{tabular}
\end{center}
 \vspace{-0.25cm}\hspace{0.5cm} a: using  1.3-in strut diameter   
\end{table} 
 
  \begin{figure}[htb]
\centering
   \includegraphics[width=.3\linewidth]{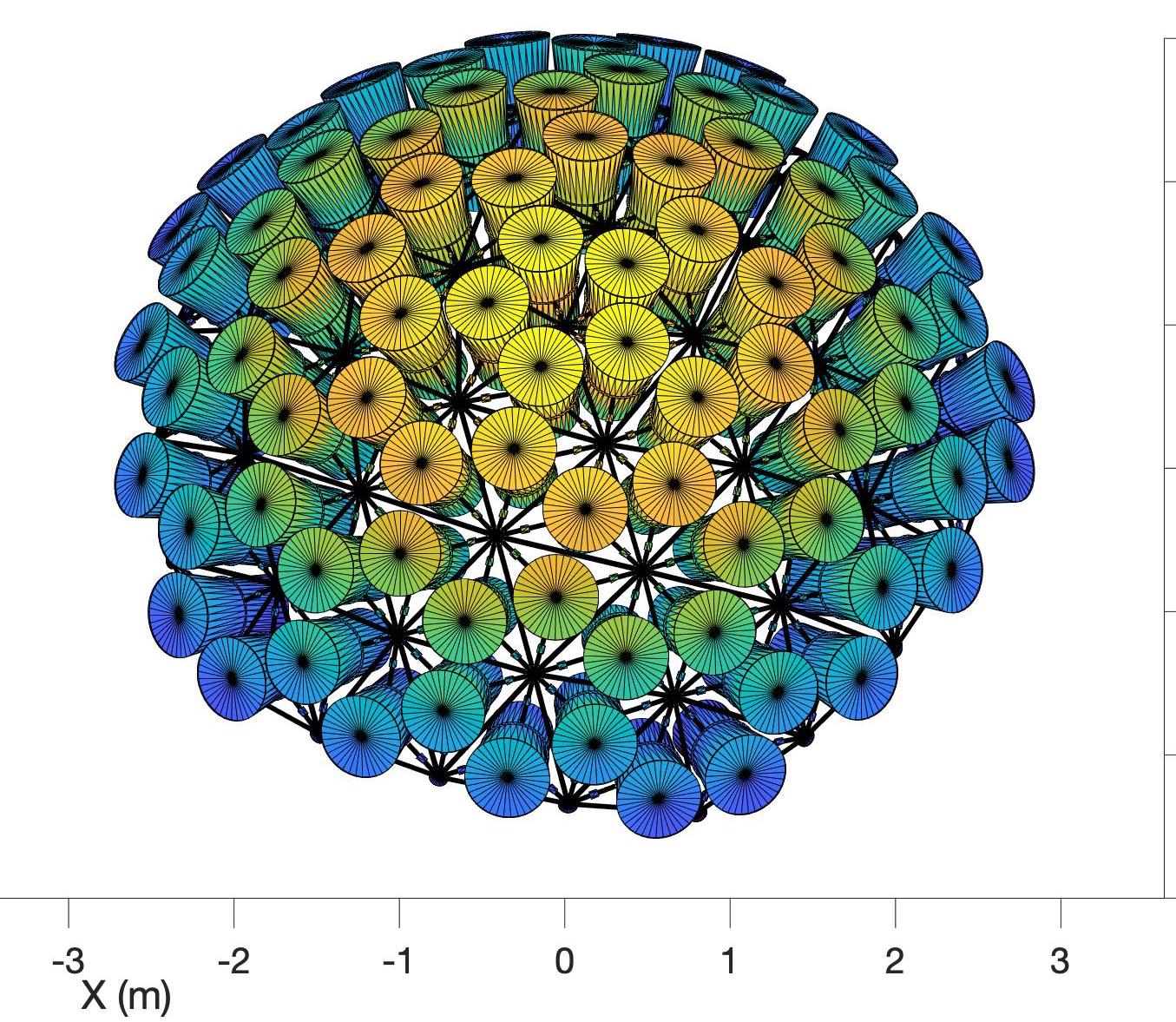}
  \includegraphics[width=.3\linewidth]{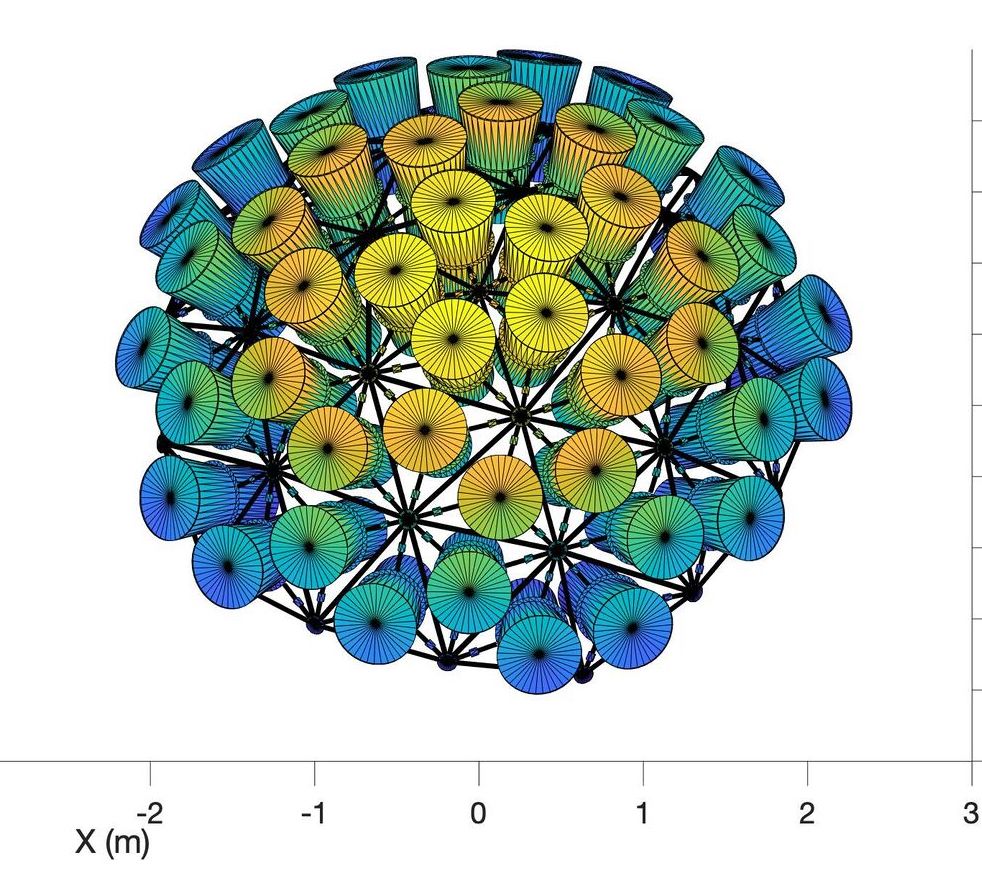}
 \includegraphics[width=.3\linewidth]{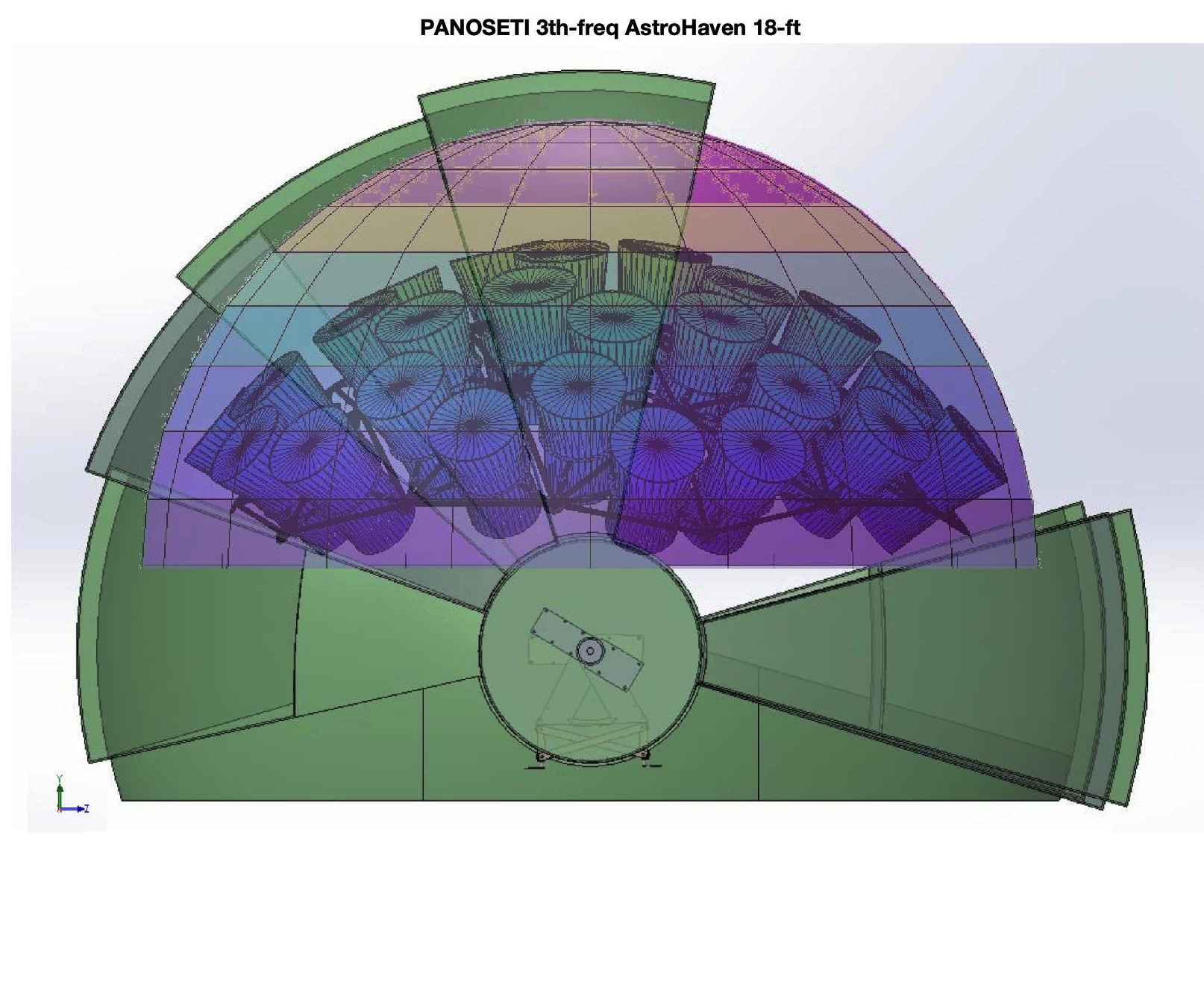}\\
    \includegraphics[width=.3\linewidth]{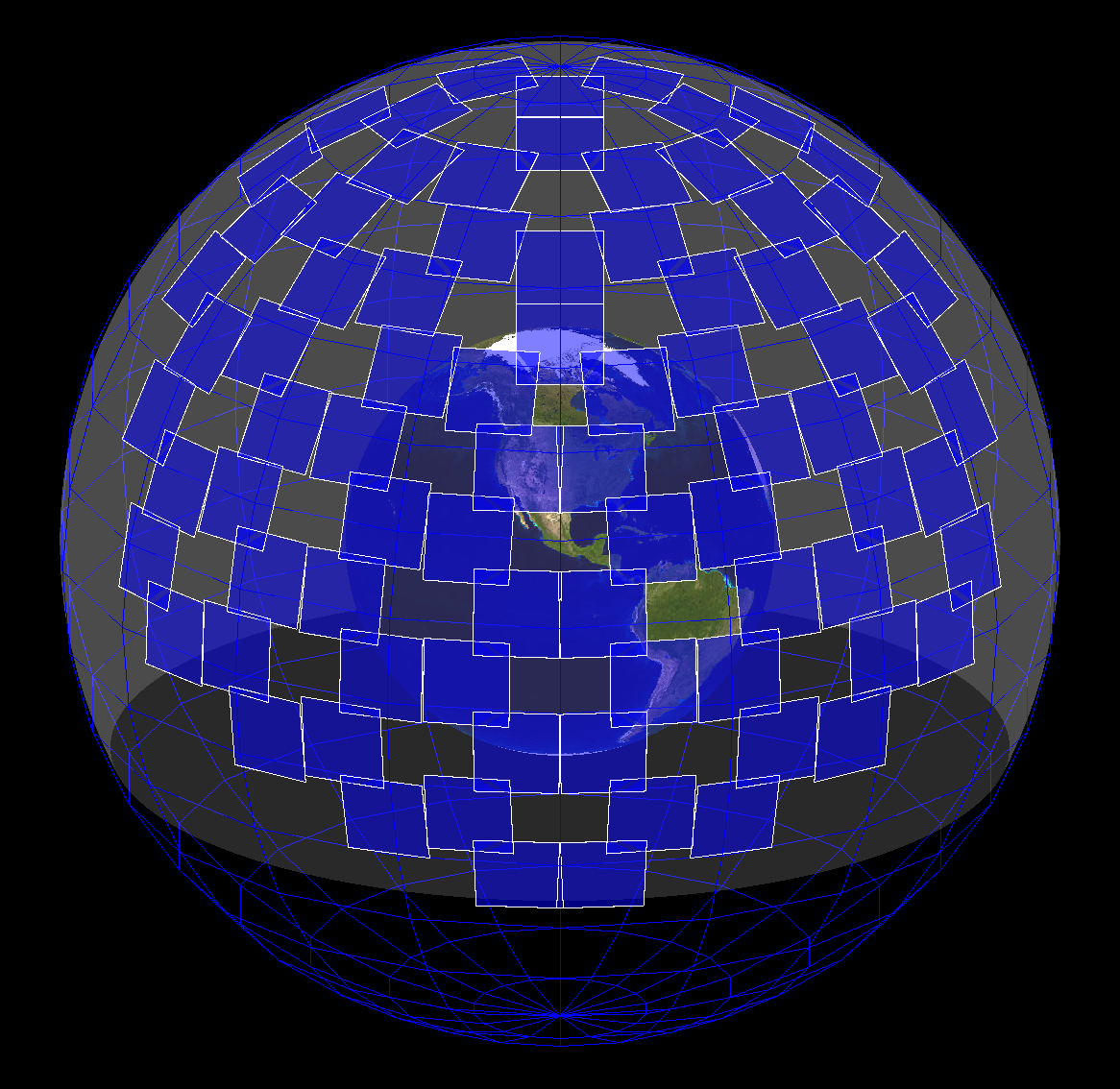}
\includegraphics[width=.31\linewidth]{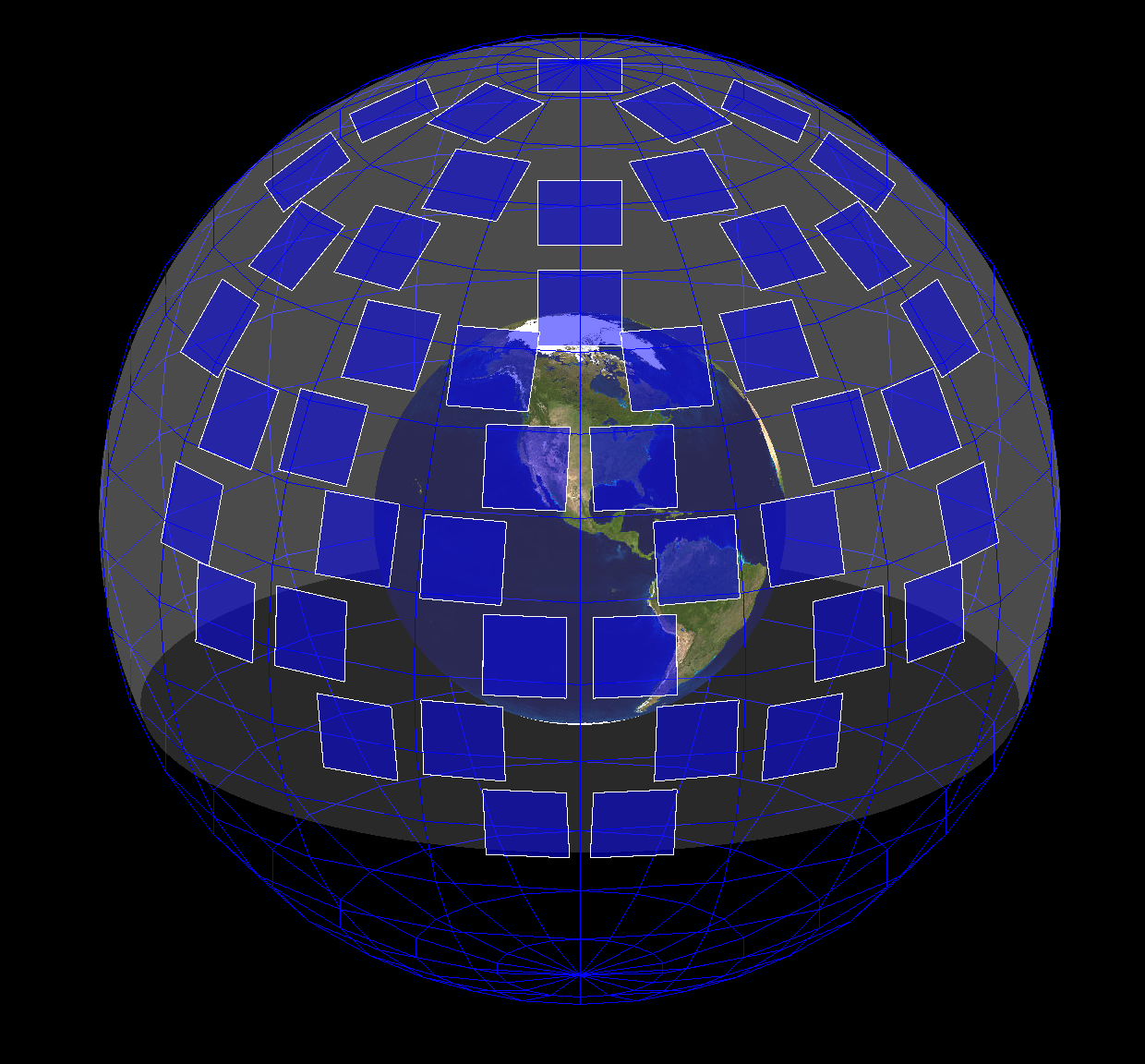}
   \includegraphics[width=.29\linewidth]{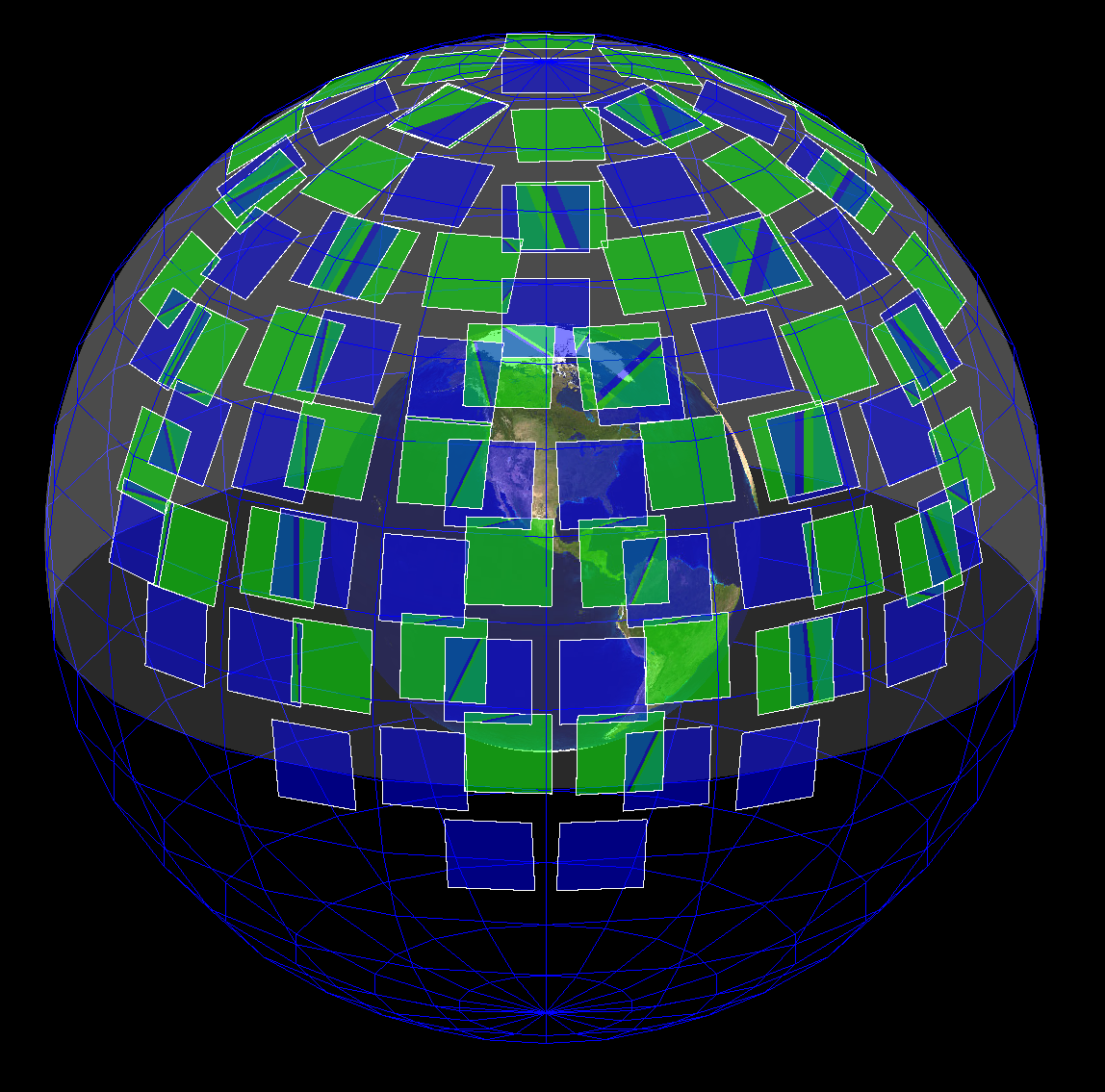}
  \caption{PANOSETI all-sky telescope assemblies and projected field-of-views in different configurations: 80 Fresnel telescopes assembly using a 4$^{th}$-frequency geodesic dome layout (top-left panel), and a configuration of 45 telescopes using a 3rd-frequency geodesic layout  (top-middle panel). Top-right panel illustrates the clearance space of the 45-telescopes configuration inside a 18-ft diameter dome enclosure. Bottom panels represents the respective instantaneous field-of-view coverage (blue squares) with an additional configuration (bottom-right panel) of two 3rd-frequency domes tilted by 14.4$^\circ$ and rotated with respect to each other (blue and green squares in field-of-views).}
  \label{fig:sub2}
\end{figure}

Table \ref{tab:onskymapping} summarizes and compares PANOSETI main assembly parameters in three possible configurations: a 4$^{th}$-frequency assembly of 80 telescopes resulting in 7,450\,sq.deg.\ of instantaneous field-of-view coverage with 442\,sq.deg.\ of redundant coverage, a  3$^{rd}$-frequency  assembly of 45 telescopes (4,441\,sq.deg.\  instantaneous field-of-view coverage without redundancy), and a double 3rd frequency assembly of 90 telescopes separated in two separate domes (7,502\,sq.deg.\ instantaneous field-of-view coverage with 19.2$\%$ redundancy). These assemblies are illustrated in Fig.\,\ref{fig:sub2} along with a representation of  their instantaneous field-of-view coverage. The configuration B using 45 telescopes for a total of 4,441\,sq.deg.\ of instantaneous field-of-view has been chosen for the PANOSETI dual observatory, given that its smaller number of telescopes allows faster deployment and ease in construction with readily available enclosures.

In summary, this PANOSETI experiment has shown proof-of-concept for detecting astronomical transients over a wide field-of-view with high sensitivity. The deployment of pairs of telescopes has confirmed the benefit of using a long baseline ($>$1\,km) to identify and characterize false alarms generated by nearby transient phenomena occurring in Earth's atmosphere. 
  


\acknowledgments 
We would like to acknowledge the Lick Observatory staff and engineers for the installation of the two PANOSETI telescopes into the Astrograph dome, with a special thanks to Adam Nichols. We would also like to thank the Palomar Observatory staff in their support of our on-site baseline observations. The PANOSETI research and instrumentation program was made possible by the support and interest of Franklin Antonio. We thank the Bloomfield Family Foundation for supporting SETI research at UC San Diego in the CASS Optical and Infrared Laboratory. Harvard SETI was supported by The Planetary Society and The Bosack/Kruger Charitable Foundation. UC Berkeley's SETI efforts involved with PANOSETI are supported by NSF grant 1407804, the Breakthrough Prize Foundation, and the Marilyn and Watson Alberts SETI Chair fund. 


\bibliography{main} 
\bibliographystyle{spiebib} 

\end{document}